\documentclass{ws}

\makeatletter
\let\trimmarks\relax
\let\cropmarks\relax

\makeatother

\usepackage{graphicx}
\usepackage{amsmath,amssymb,amsfonts}
\usepackage{newtxtext, newtxmath}

\title{NON-NORM CRITERIA AND OPTIMAL $2\times 2$ SPACE-TIME BLOCK CODES
	OVER RINGS OF INTEGERS OF IMAGINARY QUADRATIC FIELDS}

\author{CARINA ALVES\footnote{E-mails: carina.alves@unesp.br; elitonmoro@hotmail.com; cintya.benedito@unesp.br; antonio.andrade@unesp.br.}}
\address{Department of Mathematics, S\~ao Paulo State University (UNESP), Rio Claro, S\~ao Paulo, Brazil}

\author{ELITON MENDON\c{C}A MORO}
\address{S\~ao Paulo State University (UNESP), S\~ao Jos\'e do Rio Preto, S\~ao Paulo, Brazil}

\author{CINTYA WINK DE OLIVEIRA BENEDITO}
\address{School of Engineering, S\~ao Paulo State University (UNESP), S\~ao Jo\~ao da Boa Vista, S\~ao Paulo, Brazil}

\author{ANTONIO APARECIDO DE ANDRADE}
\address{Department of Mathematics, S\~ao Paulo State University (UNESP), S\~ao Jos\'e do Rio Preto, S\~ao Paulo, Brazil}

\begin{document}
	\maketitle
	
	\begin{abstract}
		Codes arising from algebraic structures over number fields lead naturally
		to determinant optimization problems governed by arithmetic invariants.
		In this paper, we investigate $2\times 2$ space-time block codes defined
		over rings of integers of imaginary quadratic fields, combining tools
		from algebraic number theory, cyclic algebras, and lattice theory.
		We prove that the Eisenstein construction over $\mathbb{Z}[\zeta_3]$ is
		optimal within the family considered here: it attains the largest
		normalized density among the $2\times 2$ space-time block codes arising
		from rings of integers of imaginary quadratic fields. As a first step, we
		show that any code that could improve upon the Eisenstein construction
		must be defined over the ring of integers of $\mathbb{Q}(\sqrt{-d})$ with
		$d\in\{2,7,11\}$, apart from the classical Gaussian and Eisenstein cases.
		We then analyze these remaining fields by explicit arithmetic arguments,
		determine the optimal constructions over them, and show that none of
		them improves upon the Eisenstein code.
		A key ingredient in our approach is the derivation of effective non-norm
		criteria for quadratic extensions of imaginary quadratic fields. These
		criteria are obtained by local methods involving $2$-adic and $3$-adic
		valuations together with Hensel's lemma, and they ensure the division
		algebra property required for full diversity. They may also be of
		independent interest in the study of division algebras and their
		applications to coding theory and lattice-based communication.
		
	\end{abstract}
	
	
	
	\section{Introduction}
	
	Algebraic constructions of space-time block codes (STBCs) from number fields and central simple algebras provide a natural framework in which determinant-type performance criteria can be studied through arithmetic invariants. In particular, cyclic division algebras and quaternion algebras yield full-diversity constructions, since the non-vanishing determinant property is governed by norm conditions and by the division algebra property \cite{20,OggierBelfioreViterbo2007}. In this setting, relevant parameters can be expressed in terms of norms, discriminants, and covolumes of associated lattices \cite{23,28}.
	
	In the $2\times 2$ case, such space-time block codes may be viewed as subsets of $\boldsymbol{M}_2(\mathbb C)$ and, after choosing integral bases, as points of a full-rank real lattice. Their determinant-based performance is commonly measured by the minimum determinant together with a normalized density, or equivalently by the normalized minimum determinant under a fixed energy normalization \cite{23,densest}. This point of view translates questions about optimality into arithmetic and geometric problems involving norm forms, field discriminants, and lattice covolumes. It also highlights the interaction between algebraic coding theory, lattice theory, and algebraic number theory \cite{28,18,index}.

	Constructions based on quadratic and cyclotomic extensions have been extensively studied, leading to full-diversity codes with favorable determinant properties \cite{FerrariAndrade2018, cyclo,wang, Wang2004, Strapasson2021}. Classical examples include the Golden code \cite{golden} and the Silver code \cite{silver}, which are obtained via division algebra techniques. Cyclic division algebras, in particular, have played a central role in the construction of full-diversity, high-rate STBCs \cite{OggierBelfioreViterbo2007, set}. Related lattice constructions arising from quaternion algebras have also been investigated over imaginary quadratic fields \cite{carina} and over totally real number fields \cite{cintya}, as well as in the context of hyperbolic lattices derived from arithmetic groups and tessellations \cite{Queiroz2016}.

	In the quadratic setting, optimal codes over the classical rings $\mathbb{Z}[i]$ and $\mathbb{Z}[\zeta_3]$ have been obtained, with the Eisenstein case providing the largest normalized density among the corresponding $2\times 2$ constructions over the Gaussian and Eisenstein integer rings \cite{cyclo,wang,golden}. This raises the question of whether this optimality persists when one considers arbitrary rings of integers of imaginary quadratic fields. 
	To address this problem, we study $2 \times 2$ STBCs defined over  the rings of integers $\mathcal{O}_{\mathbb{F}}$ of imaginary quadratic fields $\mathbb{F} = \mathbb{Q}(\sqrt{-d})$, where $d > 0$ is square-free. This framework includes the Gaussian and Eisenstein cases as particular instances, while allowing a systematic analysis over arbitrary imaginary quadratic integer rings.
	
	The study of such codes is also motivated by relatively recent  progress in algebraic reduction theory for decoding.
	While most classical constructions in space-time coding are built over the Gaussian and Eisenstein integer rings, LLL-type reduction techniques have been extended to lattices over general imaginary quadratic fields in \cite{ArimotoHirano2019, Arimoto2020}. Thus, the algorithmic tools available for decoding are no longer confined to the settings of $\mathbb{Q}(i)$ and $\mathbb{Q}(\zeta_3)$. This provides a natural incentive to study code constructions over arbitrary imaginary quadratic integer rings.

	Our main result proves a global optimality statement: the code defined over $\mathbb{Z}[\zeta_3]$ is optimal among all $2 \times 2$ space-time block codes arising from rings of integers of imaginary quadratic fields, under the normalization $\det_{\min}(\mathcal{C}) = 1$. In particular, no such code achieves a larger normalized density. 
	
	The proof is based on a determinant estimate showing that any construction which improves upon the Eisenstein code, in terms of normalized density, must have a sufficiently small covolume. Since the contribution of the base lattice associated with $\mathcal{O}_{\mathbb{F}}$ to this covolume increases with $d$, the estimate reduces the problem to the finite set $d\in\{1,2,3,7,11\}$. 
	These are precisely the values for which
	$\mathbb{Q}(\sqrt{-d})$ is norm-Euclidean among imaginary quadratic
	fields. Although this Euclidean property is not the source of the
	reduction itself, it is useful for carrying out the explicit arithmetic
	computations in the remaining cases.
	
	Excluding the classical cases $d=1$ and $d=3$, one is left with the fields $d \in \{2,7,11\}$. For each of these cases, we construct explicit STBCs and compute the corresponding determinant-based performance metrics. In particular, for $d=7$, the resulting normalized minimum determinant coincides with
	that of the Silver code \cite{silver}.

	An important tool in our approach is the derivation of effective non-norm criteria for quadratic extensions, ensuring that the associated cyclic algebras are division algebras. These criteria are obtained via local methods based on $2$-adic and $3$-adic valuations  with Hensel’s lemma. Beyond their role in the present work, these results may be of independent interest in algebraic coding theory and  in the study of division algebras over number fields.
	
	The paper is organized as follows. In Section 2, we introduce the class of $2 \times 2$ STBCs under consideration. In Section 3, we reinterpret these codes in terms of lattices and introduce a determinant-based performance metric. Section 4 establishes the determinant criterion and the reduction to bounded representatives modulo $2\mathcal{O}_{\mathbb{F}}$. Section 5 contains the main optimality results and explicit constructions for the remaining cases. Finally, Section 6 presents concluding remarks and directions for future research.

	\section{$2 \times 2$ Space-Time Block Codes from Quadratic Extensions}
	\label{sec:codes}
	
	In this section, we introduce the class of $2 \times 2$ space-time block codes considered in this paper. These codes arise from quadratic field extensions and their associated cyclic algebras.
	
	Let $\mathbb{F}$ be a number field and denote by $\mathcal{O}_{\mathbb{F}}$ its ring of integers. Fix a monic polynomial $
	x^2 + px + q \in \mathcal{O}_{\mathbb{F}}[x],
	$
	which is irreducible over $\mathbb{F}$, and let $\alpha_1, \alpha_2$ be its roots in an algebraic closure of $\mathbb{F}$. Set $\mathbb{K} = \mathbb{F}(\alpha_1)$, so that $[\mathbb{K}:\mathbb{F}] = 2$ and $\{1,\alpha_1\}$ is a basis of $\mathbb{K}$ over $\mathbb{F}$.
	
	Fix an embedding $\mathbb{F} \rightarrow \mathbb{C}$. It extends to two embeddings $\sigma_1, \sigma_2 : \mathbb{K} \rightarrow \mathbb{C}$, determined by 
	$\sigma_1(\alpha_1) = \alpha_1,$ $\sigma_2(\alpha_1) = \alpha_2.
	$
	We denote by $\mathbb{K}^\times$ the multiplicative group of nonzero elements of $\mathbb{K}$.
	
	\begin{definition} \label{qd1}
		Let $\gamma \in \mathcal{O}_{\mathbb{F}} \setminus \{0\}$  be such that $\gamma \notin N_{\mathbb{K}/\mathbb{F}}(\mathbb{K}^\times)$. The associated $2 \times 2$ space-time block code
		$
		\mathcal{C} := \mathcal{C}(\mathbb{F}, \alpha_1, \alpha_2, \gamma)
		$
		is defined as the set
		\begin{equation} \label{palavracodigo}
			\mathcal{C} =
			\left\{
			X =
			\begin{pmatrix}
				r + s\alpha_1 & t + u\alpha_1 \\
				\gamma (t + u\alpha_2) & r + s\alpha_2
			\end{pmatrix}
			: r,s,t,u \in \mathcal{O}_{\mathbb{F}}
			\right\}
			\subset \boldsymbol{M}_2(\mathbb{C}).
		\end{equation}
	\end{definition}

	Definition~\ref{qd1} admits a natural interpretation in terms of cyclic algebras. Indeed, consider the cyclic algebra
	$\mathcal{A} = (\mathbb{K}/\mathbb{F}, \sigma, \gamma),
	$
	where $\sigma$ is the nontrivial element of $\mathrm{Gal}(\mathbb{K}/\mathbb{F})$ and multiplication is defined by
	$
	e^2 = \gamma,$ $ez = \sigma(z)e,$  for all  $z \in \mathbb{K}.$
	Since $[\mathbb{K}:\mathbb{F}] = 2$, the algebra $\mathcal{A}$ is a quaternion algebra over $\mathbb{F}$. More precisely, if $\mathbb{K} = \mathbb{F}(\sqrt{a})$ for some $a \in \mathbb{F}^\times \setminus (\mathbb{F}^\times)^2$, then $\mathcal{A}$ is isomorphic to the quaternion algebra $(a,\gamma)_{\mathbb{F}}$ with basis $\{1,i,j,k\}$ satisfying
	$
	i^2 = a, $ $j^2 = \gamma,$ $ij = -ji = k.
	$
	For background on cyclic algebras and their applications to space-time coding, we refer to \cite{OggierBelfioreViterbo2007, set}. 
	
	\begin{definition} \label{defdet}
		The minimum determinant of the code $\mathcal{C} := \mathcal{C}(\mathbb{F},\alpha_1,\alpha_2,\gamma)$ is defined by
		\begin{equation} \label{eq22}
			\det_{\min}\bigl(\mathcal{C}\bigr)
			=
			\min\{\,|\det(X)| : 0\neq X\in \mathcal{C}(\mathbb{F},\alpha_1,\alpha_2,\gamma)\,\},
		\end{equation}
		where $|\cdot|$ denotes the usual complex absolute value.
	\end{definition}
	
	The following lemma establishes a fundamental property of the minimum determinant for codes defined over  the rings of integers of imaginary quadratic fields.
	
	\begin{lemma} \label{dmin1}
		Let $\mathbb{F} = \mathbb{Q}(\sqrt{-d})$, where $d > 0$ is square-free, and let
		$\mathcal{C} = \mathcal{C}(\mathbb{F},\alpha_1,\alpha_2,\gamma)$ be as in Definition~\ref{qd1}. Then
		$
		\det_{\min}(\mathcal{C}) = 1.
		$
	\end{lemma}
	
	\begin{proof}
		Let
		$
		X =
		\begin{pmatrix}
			r + s\alpha_1 & t + u\alpha_1 \\
			\gamma(t + u\alpha_2) & r + s\alpha_2
		\end{pmatrix}
		\in \mathcal{C},
		\quad r,s,t,u \in \mathcal{O}_{\mathbb{F}},
		$
		and set $g = r + s\alpha_1$, $h = t + u\alpha_1 \in \mathbb{K}$. Then
		$
		\det(X) = N_{\mathbb{K}/\mathbb{F}}(g) - \gamma N_{\mathbb{K}/\mathbb{F}}(h).
		$
		If $\det(X)=0$ and $h \neq 0$, then
		$
		\gamma = N_{\mathbb{K}/\mathbb{F}}(g/h),
		$
		which contradicts the assumption that $\gamma \notin N_{\mathbb{K}/\mathbb{F}}(\mathbb{K}^\times)$. If $h=0$, then $X \neq 0$ implies $g \neq 0$, and hence $\det(X) = N_{\mathbb{K}/\mathbb{F}}(g) \neq 0$. Therefore, $\det(X) \neq 0$ for all $X \neq 0$.
		
		Since $\alpha_1$ is a root of a monic polynomial in $\mathcal{O}_{\mathbb{F}}[x]$, it is integral over $\mathcal{O}_{\mathbb{F}}$. Thus $g=r+s\alpha_1$ and $h=t+u\alpha_1$ belong to $\mathcal{O}_{\mathbb{K}}$, and consequently 
		$N_{\mathbb{K}/\mathbb{F}}(g), N_{\mathbb{K}/\mathbb{F}}(h) \in \mathcal{O}_{\mathbb{F}}$, so  $\det(X) \in \mathcal{O}_{\mathbb{F}}$.
		Since $\mathbb{F}$ is an imaginary quadratic field, for any embedding into $\mathbb{C}$ we have $
		|x|^2 = |N_{\mathbb{F}/\mathbb{Q}}(x)| \ge 1
		$
		for all $x \in \mathcal{O}_{\mathbb{F}} \setminus \{0\}$. It follows that $|\det(X)| \ge 1$ for all $X \neq 0$.
		Finally, taking $r=1$ and $s=t=u=0$, we obtain $\det(X)=1$, which proves the result.
	\end{proof}

	\section{Lattice Representation and the Determinant-Based Performance Metric}

	In this section, we reinterpret the $2 \times 2$ space-time block codes introduced in Section 2 in terms of lattices. This viewpoint allows us to compare different constructions using geometric invariants, in particular the covolume of the associated lattice. We follow the standard approach in the literature on algebraic space-time coding (see, e.g., \cite{Wang2004, OggierBook}), and recall only the necessary definitions  for our purposes.
	
	\begin{definition} \label{retreal}
		Let $n \geq 1$ and let $M \in \boldsymbol{M}_n(\mathbb{R})$ be a full-rank matrix. The lattice generated by $M$ is defined as
		\[
		\Lambda_n(M) := \{ M \mathbf{z} \mid \mathbf{z} \in \mathbb{Z}^n \},
		\]
		where $\mathbf{z} = (z_1,\dots,z_n)^t \in \mathbb{Z}^n$. The set $\Lambda_n(M)$ is called an $n$-dimensional real lattice in $\mathbb{R}^n$, and $M$ is said to be a generator matrix of $\Lambda_n(M)$.
		The determinant (or covolume) of $\Lambda_n(M)$ is defined by
		$
		\det(\Lambda_n(M)) := |\det(M)|.
		$
		When no confusion arises, we write $\Lambda$ for $\Lambda_n(M)$.
	\end{definition}

	\begin{definition} \label{reticomplex}
		Let $M \in \boldsymbol{M}_2(\mathbb{R})$ be a full-rank matrix and denote by $\Lambda_2(M)$  the associated two-dimensional real lattice. Identifying $\mathbb C$ with $\mathbb R^2$, we regard the two-dimensional real lattice $\Lambda_2(M)\subset \mathbb R^2$ as a lattice in $\mathbb C$.
		An $n$-dimensional complex lattice $\Gamma_n(G)$ over $\Lambda_2(M)$ is defined by
		\[
		\Gamma_n(G) =
		\left\{
		G \mathbf{x} \;\middle|\;
		\mathbf{x} = (x_1,\dots,x_n)^t,\; x_i \in \Lambda_2(M)
		\right\},
		\]
		where $G \in \boldsymbol{M}_n(\mathbb{C})$ is a full-rank matrix, called a generator matrix of $\Gamma_n(G)$.
	\end{definition}
	
	The performance of the code is characterized by its minimum determinant as in Definition \ref{defdet},
	and by the covolume of the associated lattice $\Lambda$, which is given by $|\det(\Lambda)|$.

	Following the standard normalization used in space-time coding theory (see \cite{densest, OggierBook}),
	for a $2\times 2$ code viewed as a real lattice $\Lambda\subset \mathbb R^8$, we define the normalized density
	of $\Lambda$ by
	\begin{equation}\label{densidade}
		\rho(\Lambda) = \dfrac{\det_{\mathrm{min}}(\mathcal{C})^{4}}{|\det(\Lambda)|}.
	\end{equation} 
	Equivalently, the normalized minimum determinant of  $\Lambda$ is defined as
	$\delta(\Lambda)=\rho(\Lambda)^{\frac{1}{4}}.$
	Since all codes considered below satisfy $\det_{\min}(C)=1$, maximizing $\rho(\Lambda)$ is equivalent to minimizing $|\det(\Lambda)|$.

	We interpret the code construction in terms of complex lattices, following Definition~\ref{reticomplex}. 
	Let $\mathbb{F} = \mathbb{Q}(\sqrt{-d})$ be an imaginary quadratic field, and let $\mathcal{O}_{\mathbb{F}}$ denote its ring of integers. Fix a complex embedding $\iota : \mathbb{F} \hookrightarrow \mathbb{C}$. Via the identification $\mathbb{C} \simeq \mathbb{R}^2$, given by $z = x + iy \mapsto (x,y)$, the ring $\mathcal{O}_{\mathbb{F}}$ is identified with a lattice $\Lambda_2(M) \subset \mathbb{R}^2$.
	More precisely, the generator matrix $M$ is obtained by applying $\iota$ to an integral basis of $\mathcal{O}_{\mathbb{F}}$. If $d \equiv 1,2 \pmod{4}$, then $\mathcal{O}_{\mathbb{F}} = \mathbb{Z}[\sqrt{-d}]$ with basis $\{1,\sqrt{-d}\}$ and
	$
	M =
	\begin{pmatrix}
		1 & 0 \\
		0 & \sqrt{d}
	\end{pmatrix}.
	$
	If $d \equiv 3 \pmod{4}$, then $\mathcal{O}_{\mathbb{F}} = \mathbb{Z}\left[\frac{1+\sqrt{-d}}{2}\right]$ with basis $\left\{1,\frac{1+\sqrt{-d}}{2}\right\}$, and
	$
	M =
	\begin{pmatrix}
		1 & \frac{1}{2} \\
		0 & \frac{\sqrt{d}}{2}
	\end{pmatrix}.
	$
	
	Consider the $2\times 2$ space-time block code $\mathcal{C} = \mathcal{C}(\mathbb{F},\alpha_1,\alpha_2,\gamma)$ introduced in Definition~2.1. For a codeword $X \in \mathcal{C}$, define
	$
	g = r + s\alpha_1$ and $ h = t + u\alpha_1,$ where $r,s,t,u \in \mathcal{O}_\mathbb{F}$.
	
	Let $\sigma$ denote the nontrivial automorphism of $\mathrm{Gal}(\mathbb{K}/\mathbb{F})$ introduced in Section \ref{sec:codes}, so that $\sigma(\alpha_1)=\alpha_2.$
	We associate to $X$ the vectors 
	$\mathbf{v}_1 = (g, \sigma(g))^t$ and $\mathbf{v}_2 = (h, \gamma \sigma(h))^t,$
	which lie in $\mathbb{C}^2$.
	These vectors can be written as
	$
	\mathbf{v}_1 = G_1 (r,s)^t$ and 
	$\mathbf{v}_2 = G_2(t,u)^t$
	where
	\[
	G_1 =
	\begin{pmatrix}
		1 & \alpha_1 \\
		1 & \alpha_2
	\end{pmatrix}
	\qquad \mbox{and} \qquad
	G_2 =
	\begin{pmatrix}
		1 & \alpha_1 \\
		\gamma & \gamma \alpha_2
	\end{pmatrix},
	\qquad \mbox{with} \ 
	G_1,G_2 \in \boldsymbol{M}_2(\mathbb{C}).
	\]
	Since $r,s,t,u \in \mathcal{O}_\mathbb{F}$, the pairs $(r,s)$ and $(t,u)$ can be viewed as elements of $\Lambda_2(M)^2$. Hence, by Definition~\ref{reticomplex}, we have
	$
	\mathbf{v}_1 \in \Gamma_2(G_1)$ and $\mathbf{v}_2 \in \Gamma_2(G_2).
	$
	
	We now pass from complex to real lattices. For any matrix $G \in \boldsymbol{M}_2(\mathbb{C})$, write $G = A + iB$ with $A,B \in \boldsymbol{M}_2(\mathbb{R})$. Its real representation is defined by
	\[
	\widetilde{G} =
	\begin{pmatrix}
		A & -B \\
		B & A
	\end{pmatrix}
	\in \boldsymbol{M}_4(\mathbb{R}),
	\]
	which satisfies
	$
	|\det(\widetilde{G})| = |\det(G)|^2.
	$
	
	Since $\Lambda_2(M)^2$ is a lattice in $\mathbb{R}^4$ with generator matrix $\mathrm{diag}(M,M)$, the corresponding real lattices have generator matrices
	\[\widetilde{G}_1 \cdot \mathrm{diag}(M,M)
	\qquad \mbox{and} \qquad \widetilde{G}_2 \cdot \mathrm{diag}(M,M).\]
	Identifying $M_2(\mathbb{C})$ with $\mathbb{C}^4$ by vectorizing the
	entries of a codeword, and then identifying $\mathbb{C}^4$ with
	$\mathbb{R}^8$, the code $\mathcal{C}$ determines a full-rank real
	lattice $\Lambda \subset \mathbb{R}^8$. Up to a permutation of
	coordinates, a generator matrix $\mathcal{G}$  of $\Lambda$ is
	\[
	\mathcal{G} =
	\mathrm{diag}(\widetilde{G}_1 \cdot \mathrm{diag}(M,M),
	\widetilde{G}_2 \cdot \mathrm{diag}(M,M)).
	\]
	
	By multiplicativity of the determinant and since
	$|\det(\widetilde{G})| = |\det(G)|^2$, we obtain
	\[
	|\det(\mathcal{G})|
	=
	|\det(G_1)|^2\,|\det(G_2)|^2\,|\det(M)|^4.
	\]
	
	Finally, computing the determinants of $G_1$ and $G_2$, we obtain
	\[
	|\det(\mathcal{G})|
	=
	|\gamma|^2\,|\alpha_1 - \alpha_2|^4\,|\det(M)|^4.
	\]

	By Lemma \ref{dmin1} and equation~(\ref{densidade}), the normalized density can be rewritten as
	\begin{equation}\label{densidadenorm}
		\rho(\Lambda)
		=
		\frac{1}
		{|\gamma|^2\, |\alpha_1 - \alpha_2|^4\, |\det(M)|^4}.
	\end{equation}
	
	Maximizing the normalized density is equivalent to minimizing the denominator of (\ref{densidadenorm}). This motivates the following definition.
	
	\begin{definition}\label{basedmetric}
		Let $\mathcal{C}(\mathbb{F},\alpha_1,\alpha_2,\gamma)$ be a space-time block code. The determinant-based metric of $\mathcal{C}$ is defined by
		\begin{equation} \label{cdet}
			c_{\det}(\mathcal{C}(\mathbb{F},\alpha_1,\alpha_2,\gamma)) = |\gamma| \, |\alpha_1 - \alpha_2|^2 \, |\det(M)|^2.
		\end{equation}
	\end{definition}
	
	This quantity will be used throughout the paper to compare different code constructions.

	\section{Determinant Criterion and Parameter Reduction over $\mathcal{O}_{\mathbb{F}}$}
	
	Using the framework developed in Section~3, we adopt the determinant-based comparison criterion introduced in \cite{cyclo} to compare $2 \times 2$ space-time block codes. The comparison is based on the determinant-based metric defined in Definition~\ref{basedmetric}.

	\begin{definition}[Determinant criterion]
		Let
		\[
		\mathcal{C}_1=\mathcal{C}(\mathbb{F}_1,\alpha_1,\alpha_2,\gamma_1)
		\qquad\text{and}\qquad
		\mathcal{C}_2=\mathcal{C}(\mathbb{F}_2,\beta_1,\beta_2,\gamma_2)
		\]
		be two $2\times 2$ space-time block codes such that
		$
		\det_{\mathrm{min}}(\mathcal{C}_1)=\det_{\min}(\mathcal{C}_2).
		$
		We say that $\mathcal{C}_1$ is better than $\mathcal{C}_2$ if the determinant-based metric
		$
		c_{\det}(\mathcal{C}_1)<c_{\det}(\mathcal{C}_2).
		$
		Equivalently, this means that the normalized density of the lattice associated with $\mathcal{C}_1$ is larger than that of the lattice associated with $\mathcal{C}_2$.
	\end{definition}
	
	\begin{definition} \label{S}
		Let $\mathcal{S}$ be a family of $2\times 2$ space-time block codes, all having the same minimum determinant. A code $C\in \mathcal{S}$ is called optimal in $\mathcal{S}$ if
		\[
		c_{\det}(C)\leq c_{\det}(\widetilde{C})
		\quad\text{for every } \widetilde{C}\in \mathcal{S}.
		\]
		
	\end{definition}

	The following lemma generalizes a property observed in \cite{wang}, where two specific cases
	are considered.

	\begin{lemma} \label{transla}
		Let $\mathbb{F}$ be an imaginary quadratic field with ring of integers $\mathcal{O}_{\mathbb{F}}$,
		and let $\mathcal{C} = \mathcal{C}(\mathbb{F},\alpha_1,\alpha_2,\gamma)$ be a space-time block code as in Section~2.
		Let $p_0 \in \mathcal{O}_{\mathbb{F}}$ and define
		$\alpha_1'=\alpha_1+p_0$ and $\alpha_2'=\alpha_2+p_0.$
		Then
		\[
		c_{\det}\bigl(\mathcal{C}(\mathbb{F},\alpha_1',\alpha_2',\gamma)\bigr)
		=
		c_{\det}\bigl(\mathcal{C}(\mathbb{F},\alpha_1,\alpha_2,\gamma)\bigr).
		\]
		Moreover, if
		$
		f(x)=x^2+px+q \in \mathcal{O}_{\mathbb{F}}[x]
		$
		is the minimal polynomial of $\alpha_1$ over $\mathbb{F}$, then the minimal polynomial of $\alpha_1'$ over $\mathbb{F}$ is
		\[
		f_{p_0}(x)=x^2+(p-2p_0)x+(q-pp_0+p_0^2),
		\]
		which has $\alpha_1'$ and $\alpha_2'$ as roots.
	\end{lemma}
	
	\begin{proof}
		Since
		$
		\alpha_1'-\alpha_2'=\alpha_1-\alpha_2,
		$
		it follows directly from Definition \ref{basedmetric} that
		\[
		c_{\det}\bigl(\mathcal{C}(\mathbb{F},\alpha_1',\alpha_2',\gamma)\bigr)
		=
		c_{\det}\bigl(\mathcal{C}(\mathbb{F},\alpha_1,\alpha_2,\gamma)\bigr).
		\]
		If
		$
		f(x)=x^2+px+q
		$
		is the minimal polynomial of $\alpha_1$ over $\mathbb{F}$, then the minimal polynomial of $\alpha_1'$ is obtained from $f$ by the change of variable $x\mapsto x-p_0$, namely
		$
		f_{p_0}(x)=f(x-p_0).
		$
		
		A direct computation yields
		\[
		f_{p_0}(x)=x^2+(p-2p_0)x+(q-pp_0+p_0^2),
		\]
		which completes the proof.
		
	\end{proof}
	
	By Lemma \ref{transla}, replacing $\alpha_i$ with $\alpha_i+p_0$, for $i=1,2$, does not change the difference $\alpha_1-\alpha_2$ and therefore leaves the determinant-based metric unchanged. Moreover, this change of variable preserves irreducibility and yields an isomorphic quadratic extension. Hence, for the purpose of minimizing $c_{\det}$, the coefficient $p$ may be replaced by any representative of its class modulo $2\mathcal{O}_{\mathbb{F}}$
	without affecting 
	$c_{\det}$. In particular, it is natural to choose $p_0$ such that
	$
	p'=p-2p_0
	$
	it has a small absolute value. The next proposition shows that such a choice can be made with a uniform bound depending only on $d$. We refer to this as parameter reduction over $\mathcal{O}_{\mathbb{F}}$.
	
	\begin{proposition} \label{cases}
		Let
		$
		\mathbb{F}=\mathbb{Q}(\sqrt{-d}),
		$
		where $d>0$ is square-free, and let $\mathcal{O}_{\mathbb{F}}$ be its ring of integers.
		Fix a complex embedding of\,      $\mathbb{F}$ into $\mathbb{C}$
		and denote by $|\cdot|$ the corresponding complex modulus.
		Then, for every $p\in \mathcal{O}_{\mathbb{F}}$, there exists $p_0\in \mathcal{O}_{\mathbb{F}}$ such that
		\[
		|p-2p_0|\leq
		\begin{cases}
			\max\left\{1,\left|\dfrac{1+\sqrt{-d}}{2}\right|\right\}, & \text{if } d\equiv 3 \pmod{4}\\[10pt]
			|1+\sqrt{-d}|, & \text{if } d\equiv 1,2 \pmod{4}
		\end{cases}.
		\]
	\end{proposition}
	\begin{proof} 
		We first suppose that $d \equiv 3 \pmod{4}$. Then $\mathcal{O}_{\mathbb{F}} = \mathbb{Z}[\omega]$, where $\omega = \frac{1+\sqrt{-d}}{2}$, so every $p \in \mathcal{O}_{\mathbb{F}}$ can be written in the form $p = a + b\omega$, with $a,b \in \mathbb{Z}$. 
		Choose $r,s \in \mathbb{Z}$ such that $a - 2r \in \{0,-1\}$ and $b - 2s \in \{0,1\}$, which is always possible by reduction modulo $2$. Setting $p_0 = r + s\omega$, we obtain
		$
		p - 2p_0 = (a - 2r) + (b - 2s)\omega,
		$
		which lies in the set $\{0,\omega,-1,-1+\omega\}$. Hence,
		$
		|p - 2p_0| \leq \max\{0,|\omega|,1,|-1+\omega|\}.
		$
		Since $|-1+\omega| = |\omega|$, it follows that
		$
		|p - 2p_0| \leq \max\{1,|\omega|\} = \max\left\{1,\left|\frac{1+\sqrt{-d}}{2}\right|\right\}.
		$
		
		We now suppose that $d \equiv 1,2 \pmod{4}$. Then $\mathcal{O}_{\mathbb{F}} = \mathbb{Z}[\sqrt{-d}]$, and every $p \in \mathcal{O}_{\mathbb{F}}$ can be written in the form $p = a + b\sqrt{-d}$, with $a,b \in \mathbb{Z}$. 
		Choose $r,s \in \mathbb{Z}$ such that $a - 2r \in \{0,1\}$ and $b - 2s \in \{0,1\}$, again possible by reduction modulo $2$. Setting $p_0 = r + s\sqrt{-d}$, we obtain
		$
		p - 2p_0 = (a - 2r) + (b - 2s)\sqrt{-d},
		$
		which lies in the set $\{0,1,\sqrt{-d},1+\sqrt{-d}\}$. Consequently,
		$
		|p - 2p_0| \leq \max\{0,1,\sqrt{d},|1+\sqrt{-d}|\}.
		$
		Since $|1+\sqrt{-d}| = \sqrt{1+d} \geq \max\{1,\sqrt{d}\}$, it follows that
		$
		|p - 2p_0| \leq |1+\sqrt{-d}|.
		$ This completes the proof.
	\end{proof}

	\section{Optimal $2\times 2$ Space-Time Block Codes over Imaginary Quadratic Integer Rings} \label{sec5}
	
	In this section, we determine optimal $2 \times 2$ space-time block codes arising from the rings of integers of imaginary quadratic fields with respect to the normalized density, under the normalization $\det_{\min}(\mathcal{C}) = 1$. In particular, we identify the optimal code over each field corresponding to $d \in \{2,7, 11\}$ and show that the Eisenstein construction over $\mathbb{Q}(\sqrt{-3})$ is globally optimal.
	
	The Eisenstein construction over $\mathbb{F} = \mathbb{Q}(\sqrt{-3})$ provides a natural reference point. As shown in \cite{wang}, the code $\mathcal{C}_{\mathbb{Q}(\sqrt{-3})}$ is optimal over $\mathcal{O}_{\mathbb{F}} = \mathbb{Z}[\zeta_3]$, with determinant-based metric 
	$
	c_{\det}(\mathcal{C}_{\mathbb{Q}(\sqrt{-3})}) =|\zeta_6||(1+\zeta_6)^2-4\sqrt{-3}|\frac{3}{4}=\frac{3}{4}\sqrt{21} \approx 3.44,
	$
	corresponding to normalized density $\rho \approx 0.085$, which exceeds the corresponding value for the Golden code over $\mathbb{Q}(i)$ \cite{golden}.
	
	Let $\mathcal{C} = \mathcal{C}(\mathbb{F}, \beta_1, \beta_2, \gamma)$, where $\mathbb{F} = \mathbb{Q}(\sqrt{-d})$, and assume that
	$
	c_{\det}(\mathcal{C}) < c_{\det}(\mathcal{C}_{\mathbb{Q}(\sqrt{-3})})
	$, that is,
	$ |\gamma||\beta_1-\beta_2|^2|\det(M)|^2 <\frac{3\sqrt{21}}{4}$.  
	
	Since $(\beta_1-\beta_2)^2\in\mathcal{O}_{\mathbb{F}}\setminus\{0\}$ and $\mathbb{F}$ is an imaginary quadratic field, we have
	$|\beta_1-\beta_2|^2=|(\beta_1-\beta_2)^2|\ge 1.
	$
	Similarly, $|\gamma|\ge 1$ for every nonzero $\gamma\in\mathcal{O}_{\mathbb{F}}$.
	It follows that
	$|\det(M)|^2 < \frac{3\sqrt{21}}{4}$.

	The quantity $|\det(M)|^2$ depends only on the residue class of $d$ modulo $4$ and satisfies
	\[
	|\det(M)|^2 =
	\begin{cases}
		\frac{d}{4}, & \text{if } d \equiv 3 \pmod{4}, \\
		d, & \text{if } d \equiv 1,2 \pmod{4}.
	\end{cases}
	\]
	
	Combining these inequalities, we obtain
	$d \in \{1,2,3,7,11\},$
	thus reducing the problem to a finite set of cases.
	
	The case $d=1$ corresponds to the Gaussian construction. Since the normalized density of the Eisenstein code exceeds that of the Gaussian code, any code with larger normalized density must arise from the cases $d \in \{2,7,11\}$.
	For each of these values, we determine admissible parameters $\alpha_1,\alpha_2,\gamma$ satisfying the non-norm condition and evaluate the corresponding determinant-based metric.
	
	The following propositions establish the non-norm criteria required for this analysis.
	
	\begin{proposition}\label{-1padic} Let $d>0$ be such that $-d\equiv 1 \pmod{3}$, and set $ \mathbb{F}=\mathbb{Q}(\sqrt{-d}),$ $\mathbb{K}=\mathbb{Q}(\sqrt{-d},\sqrt{-3}). $ Then $-1$ is not a norm from $\mathbb{K}$ to $\mathbb{F}$. \end{proposition}
	\begin{proof}
		We work locally at the prime $3$ in the field $\mathbb{Q}_3$. Since $-d \equiv 1 \pmod 3$, the polynomial $f(x)=x^2+d$ satisfies
		$f(1)=1+d \equiv 0 \pmod 3$ and $f'(1)=2 \not\equiv 0 \pmod 3$.
		By Hensel's lemma, $f$ has a root in $\mathbb{Z}_3$. Hence, $-d$ is a square in $\mathbb{Q}_3$, and therefore
		$\mathbb{F}=\mathbb{Q}(\sqrt{-d})$ admits an embedding into $\mathbb{Q}_3$. Fix such an embedding. It then follows that
		$\mathbb{K}=\mathbb{Q}(\sqrt{-d},\sqrt{-3})$ embeds into $\mathbb{Q}_3(\sqrt{-3})$.
		
		Assume, for contradiction, that $-1$ is a norm from $\mathbb{K}$ to $\mathbb{F}$. Then its image in $\mathbb{Q}_3$ would also be a norm from $\mathbb{Q}_3(\sqrt{-3})$ to $\mathbb{Q}_3$. Thus, it suffices to show that
		$-1 \notin N_{\mathbb{Q}_3(\sqrt{-3})/\mathbb{Q}_3}(\mathbb{Q}_3(\sqrt{-3})^{\times})$.
		
		Let $a,b \in \mathbb{Q}_3$. Since $N(a+b\sqrt{-3})=a^2+3b^2$, it suffices to show that the equation
		\begin{equation}\label{valo3}
			a^2+3b^2=-1
		\end{equation}
		has no solution in $\mathbb{Q}_3$.
		
		Let $v_3$ denote the $3$-adic valuation. We first show that any such solution must satisfy $a,b \in \mathbb{Z}_3$. Suppose that at least one of $v_3(a)$ or $v_3(b)$ is negative. Since $v_3(a^2+3b^2)=v_3(-1)=0$, the valuations of $a^2$ and $3b^2$ cannot be distinct; otherwise, by the ultrametric inequality, we would have
		$v_3(a^2+3b^2)=\min\{v_3(a^2),v_3(3b^2)\}<0$, a contradiction. Hence $v_3(a^2)=v_3(3b^2)$, that is, $2v_3(a)=1+2v_3(b)$, which is impossible, since the left-hand side is even whereas the right-hand side is odd. Therefore, $v_3(a)\ge 0$ and $v_3(b)\ge 0$, so $a,b \in \mathbb{Z}_3$.
		
		Reducing \eqref{valo3} modulo $3$, that is, in the residue field $\mathbb{F}_3$, we obtain
		$a^2 \equiv -1 \equiv 2 \pmod{3}.
		$
		Since the only quadratic residues modulo $3$ are $0$ and $1$, this is impossible. Hence, \eqref{valo3} has no solution in $\mathbb{Q}_3$.
		
		Consequently, $-1 \notin N_{\mathbb{Q}_3(\sqrt{-3})/\mathbb{Q}_3}\big(\mathbb{Q}_3(\sqrt{-3})^\times\big)$, and therefore $-1$ is not a norm from $\mathbb{K}$ to $\mathbb{F}$.
	\end{proof}

	\begin{proposition}\label{-1_2adic}
		Let $d>0$ be such that $-d\equiv 1 \pmod{8}$, and set
		$
		\mathbb{F}=\mathbb{Q}(\sqrt{-d}), 
		\mathbb{K}=\mathbb{Q}(\sqrt{-d},i).
		$
		Then $-1$ is not a norm from $\mathbb{K}$ to $\mathbb{F}$.
	\end{proposition}
	
	\begin{proof}
		We work locally at the prime $2$  in the field $\mathbb{Q}_2$. Let $f(x)=x^2+d$. Since $-d \equiv 1 \pmod 8$, we have $f(1)=1+d \equiv 0 \pmod 8$ and $f'(1)=2$. Since
		$
		v_2(f(1)) \ge 3 > 2 = 2v_2(f'(1)),
		$
		Hensel's lemma implies that $f$ has a root in $\mathbb{Z}_2$. Therefore $-d$ is a square in $\mathbb{Q}_2$, and so $\mathbb{F}=\mathbb{Q}(\sqrt{-d})$ embeds into $\mathbb{Q}_2$. Consequently, $\mathbb{K}=\mathbb{Q}(\sqrt{-d},i)$ embeds into $\mathbb{Q}_2(i)$.
		
		Assume, for contradiction, that $-1$ is a norm from $\mathbb{K}$ to $\mathbb{F}$. Then its image under the embedding $\mathbb{F}\hookrightarrow \mathbb{Q}_2$ would also be a norm from $\mathbb{Q}_2(i)$ to $\mathbb{Q}_2$. Thus, it suffices to show that
		$
		-1 \notin N_{\mathbb{Q}_2(i)/\mathbb{Q}_2}(\mathbb{Q}_2(i)^{\times}).
		$
		
		Let $a,b \in \mathbb{Q}_2$. Since $N(a+bi)=a^2+b^2$, it suffices to show that the equation
		\begin{equation}\label{valo4}
			a^2+b^2=-1
		\end{equation}
		has no solution in $\mathbb{Q}_2$.
		
		Let $v_2$ denote the $2$-adic valuation. We first show that any such solution must satisfy $a,b \in \mathbb{Z}_2$. Suppose that at least one of $v_2(a)$ or $v_2(b)$ is negative, and set $m=\min\{v_2(a),v_2(b)\}<0$. Write $a=2^m u$ and $b=2^m v$ with $u,v \in \mathbb{Z}_2$, at least one of which is a $2$-adic unit. Then
		\[
		a^2+b^2 = 2^{2m}(u^2+v^2)=-1,
		\]
		so $u^2+v^2=-2^{-2m}$. Therefore $v_2(u^2+v^2)=-2m\ge 2$.
		
		We claim that $v_2(u^2+v^2)\le 1$, which yields a contradiction. If exactly one of $u$ and $v$ is a unit, then $u^2+v^2$ is odd, so $v_2(u^2+v^2)=0$. If both are units, then $u^2 \equiv v^2 \equiv 1 \pmod 8$, hence $u^2+v^2 \equiv 2 \pmod 8,$ and therefore $v_2(u^2+v^2)=1$. In either case, $v_2(u^2+v^2)\le 1$, contradicting $v_2(u^2+v^2)\ge 2$.
		Thus $a,b \in \mathbb{Z}_2$. Reducing the equation \eqref{valo4} modulo $4$, we obtain $a^2+b^2 \equiv -1 \equiv 3 \pmod 4$. Since the squares modulo $4$ are $0$ and $1$, this is impossible. Hence \eqref{valo4} has no solution in $\mathbb{Q}_2$.
		Hence $-1$ is not a norm from $\mathbb{Q}_2(i)$ to $\mathbb{Q}_2$, and thus not a norm from $\mathbb{K}$ to $\mathbb{F}$.
	\end{proof}
	
	The remaining cases $d \in \{2,7,11\}$ are analyzed in the following subsection.

	\subsection{Optimal $2\times 2$ space-time block codes over $\mathbb{Z}[\sqrt{-2}]$}\label{subsec:d2}
	
	We now treat the case $\mathbb{F}=\mathbb{Q}(\sqrt{-2})$ and determine the 
	optimal $2\times 2$ space-time block code 
	$\mathcal{C}(\mathbb{F},\alpha_1,\alpha_2,\gamma)$ over $\mathbb{Z}[\sqrt{-2}]$ in the class $\mathcal{S}$ introduced in Definition~\ref{S}.
	
	By Proposition~\ref{-1padic}, one has 
	$-1 \notin N_{\mathbb{K}/\mathbb{F}}(\mathbb{K}^{\times}),$
	where $\mathbb{F}=\mathbb{Q}(\sqrt{-2})$ and $\mathbb{K}=\mathbb{Q}(\sqrt{-2},\sqrt{-3})$. Thus, $\gamma = -1$ is an admissible choice.

	\begin{theorem}\label{thm:d2}
		The code $\mathcal{C}_{\mathbb{Q}(\sqrt{-2})}=\mathcal{C}(\mathbb{F},\alpha_1,\alpha_2,\gamma)$, with $\mathbb{F}=\mathbb{Q}(\sqrt{-2})$, $\alpha_1=\frac{1+\sqrt{-3}}{2}$, $\alpha_2=\frac{1-\sqrt{-3}}{2}$, and $\gamma=-1$, is optimal among all $2\times 2$ space-time block codes defined over $\mathbb{Z}[\sqrt{-2}]$ with minimum determinant equal to $1$.
	\end{theorem}
	
	\begin{proof}
		Let $\mathbb{F}=\mathbb{Q}(\sqrt{-2})$ and $\mathcal{O}_{\mathbb{F}}=\mathbb{Z}[\sqrt{-2}]$. By Lemma~\ref{dmin1}, all codes under consideration satisfy $\det_{\min}(\mathcal{C})=1$, so optimality is determined by the determinant-based metric $c_{\det}$.
		For the code $\mathcal{C}_{\mathbb{Q}(\sqrt{-2})}$, we have $\alpha_1-\alpha_2=\sqrt{-3}$ and $\gamma=-1$, hence $|\alpha_1-\alpha_2|^2=3$ and $|\gamma|=1$. Since $2\equiv 2 \pmod{4}$, it follows that $|\det(M)|^2=2$. Therefore,
		\[
		c_{\det}\bigl(\mathcal{C}_{\mathbb{Q}(\sqrt{-2})}\bigr)
		=
		|\gamma|\,|\alpha_1-\alpha_2|^2\,|\det(M)|^2
		=
		6.
		\]

		Assume, for contradiction, that there exists another code $\widetilde{\mathcal{C}}=\mathcal{C}(\mathbb{F},\beta_1,\beta_2,\gamma)$ over $\mathcal{O}_{\mathbb{F}}$ such that $\det_{\min}(\widetilde{\mathcal{C}})=1$ and $c_{\det}(\widetilde{ \mathcal{C}})<6$.

		Let $x^2+px+q \in \mathcal{O}_{\mathbb{F}}[x]$ be irreducible over $\mathbb{F}$, and let $\beta_1,\beta_2$ denote its roots, where $p,q \in \mathcal{O}_{\mathbb{F}}$. Then $|\beta_1-\beta_2|^2 = |p^2 - 4q|$. Since $|\det(M)|^2 = 2$, the condition $c_{\det}(\widetilde{\mathcal{C}}) < 6$ is equivalent to
		\begin{equation}\label{eq:key}
			|\gamma|\,|p^2 - 4q| < 3.
		\end{equation}
		
		By Proposition~\ref{cases}, we may assume $|p|\le |1+\sqrt{-2}|=\sqrt{3}$. Writing $p=a+b\sqrt{-2}$ with $a,b\in\mathbb{Z}$, this yields $a^2+2b^2\le 3$, and hence
		$
		p\in\{0,\pm1,\pm\sqrt{-2},\pm(1+\sqrt{-2}),\pm(1-\sqrt{-2})\}.
		$
		\begin{itemize}
			\item[\emph{Case 1:}] $p\in\{0,\pm1\}$.
			Then $p^2\in\{0,1\}$ and, since $q\neq 0$, we have $|q|\ge 1$. Hence $|p^2-4q|\ge 3$. Since $\gamma\in\mathcal{O}_{\mathbb{F}}\setminus\{0\}$ implies $|\gamma|\ge 1$, it follows that $|\gamma|\,|p^2-4q|\ge 3$, contradicting \eqref{eq:key}.
			\item[\emph{Case 2:}] $p=\pm\sqrt{-2}$.
			Then $p^2=-2$, and \eqref{eq:key} becomes $|\gamma|\,|-2-4q|<3$. Writing $q=r+s\sqrt{-2}$ with $r,s\in\mathbb{Z}$, then \[|-2-4q|^2=(2+4r)^2+32s^2.\] If $s\neq 0$, then $(2+4r)^2\ge 4$ and $32s^2\ge 32$, so $|-2-4q|\ge 6$, contradicting \eqref{eq:key}. Thus $s=0$, and $|-2-4q|=|2+4r|$. The minimal nonzero value is $2$, attained for $r=-1$, that is, $q=-1$. Hence $|p^2-4q|=2$, and \eqref{eq:key} implies $|\gamma|<\frac{3}{2}$.
			Write $\gamma = x + y\sqrt{-2}$ with $x,y\in\mathbb{Z}$. Then
			$
			|\gamma|^2 = x^2 + 2y^2.
			$
			If $|\gamma| < \frac{3}{2}$, then
			$
			x^2 + 2y^2 < \frac{9}{4}.
			$
			It follows that
			$
			x^2 + 2y^2 \in \{0,1,2\},
			$
			since $x^2 + 2y^2$ is a nonnegative integer. Hence $(x,y)\in\{(0,0),(\pm1,0),(0,\pm1)\}$, and therefore
			$
			\gamma \in \{0,\pm1,\pm\sqrt{-2}\}.
			$
			
			For $p=\pm\sqrt{-2}$ and $q=-1$, one has $p^2-4q=2,$ and therefore  $\tilde{\mathbb{K}}=\mathbb{F}(\sqrt{2})=\mathbb{Q}(\sqrt{-2},\sqrt{2})$, and
			\[
			\begin{alignedat}{2}
				&N_{\widetilde{\mathbb K}/\mathbb F}(1)=1,
				\qquad&
				&N_{\widetilde{\mathbb K}/\mathbb F}(1+\sqrt{2})=-1,\\
				&N_{\widetilde{\mathbb K}/\mathbb F}\left(-1-\frac{\sqrt{-2}}{2}+\frac{\sqrt{2}}{2}\right)=\sqrt{-2},
				\qquad&
				&N_{\widetilde{\mathbb K}/\mathbb F}\left(-1+\frac{\sqrt{-2}}{2}+\frac{\sqrt{2}}{2}\right)=-\sqrt{-2}.
			\end{alignedat}
			\]
			Hence every nonzero element in $\{\pm1,\pm\sqrt{-2}\}$ belongs to
			$N_{\tilde{\mathbb{K}}/\mathbb F}(\tilde{\mathbb{K}}^\times)$, contradicting the non-norm condition in Definition~\ref{qd1}.

			\item[\emph{Case 3:}] $p\in\{\pm(1+\sqrt{-2}),\,\pm(1-\sqrt{-2})\}$.
			Since replacing $p$ by $-p$ does not affect $p^2$, and complex conjugation preserves $|p^2-4q|$, it suffices to consider $p=1+\sqrt{-2}$. Writing $q=r+s\sqrt{-2}$, then 
			\[
			p^2-4q=(-1-4r)+(2-4s)\sqrt{-2}.
			\]
			Setting $A=-1-4r$ and $B=2-4s$, we have $|A|\ge 1$ and $|B|\ge 2$, hence $|p^2-4q|^2=A^2+2B^2\ge 9$, so $|p^2-4q|\ge 3$. Therefore, $|\gamma|\,|p^2-4q|\ge 3$, contradicting \eqref{eq:key}.
		\end{itemize}

		This contradiction excludes all possible reduced values of $p$, and therefore no admissible code can have a smaller determinant-based metric. The asserted optimality follows.
		
	\end{proof}

	\subsection{Optimal $2\times 2$ Space-Time Block Codes over $\mathbb{Z}\!\left[\frac{1+\sqrt{-7}}{2}\right]$}

	We next consider $\mathbb{F}=\mathbb{Q}(\sqrt{-7})$, whose ring of integers is $\mathbb{Z}\left[\frac{1+\sqrt{-7}}{2}\right]$ and we construct an optimal $2\times 2$ space-time block code 
	$\mathcal{C}(\mathbb{F},\alpha_1,\alpha_2,\gamma)$ over $\mathbb{Z}\!\left[\frac{1+\sqrt{-7}}{2}\right]$ 
	in the class $\mathcal{S}$ introduced in Definition~\ref{S}. 
	By Proposition~\ref{-1_2adic},
	$
	-1 \notin N_{\mathbb{K}/\mathbb{F}}(\mathbb{K}^{\times}),
	$
	where $\mathbb{F}=\mathbb{Q}(\sqrt{-7})$ and $\mathbb{K}=\mathbb{Q}(\sqrt{-7},\sqrt{-1})$. 
	Therefore, $\gamma = -1$ is admissible.
	
	\begin{theorem}\label{thm:d7}
		The code $\mathcal{C}_{\mathbb{Q}(\sqrt{-7})}=\mathcal{C}(\mathbb{F},\alpha_1,\alpha_2,\gamma)$, with $\mathbb{F}=\mathbb{Q}(\sqrt{-7})$, $\alpha_1=i$, $\alpha_2=-i$, and $\gamma=-1$, is optimal among all $2\times 2$ space–time block codes defined over $\mathbb{Z}\!\left[\frac{1+\sqrt{-7}}{2}\right]$ with minimum determinant equal to $1$.
	\end{theorem}
	\begin{proof}
		Let $\mathbb{F}=\mathbb{Q}(\sqrt{-7})$ and $\mathcal{O}_{\mathbb{F}}=\mathbb{Z}[\omega]$, where $\omega=\frac{1+\sqrt{-7}}{2}$. By Lemma~\ref{dmin1}, all codes under consideration have minimum determinant equal to $1$, so optimality is determined by the determinant-based metric $c_{\det}$.
		For the reference code $\mathcal{C}_{\mathbb{Q}(\sqrt{-7})}$, the polynomial $x^2+1$ is irreducible over $\mathbb{F}$ and therefore defines a quadratic extension. It follows that $\alpha_1=i$ and $\alpha_2=-i$, so $\alpha_1-\alpha_2=2i$ and $|\alpha_1-\alpha_2|^2=4$. Moreover, $\gamma=-1$, hence $|\gamma|=1$. Since $7\equiv 3 \pmod{4}$, the generator matrix $M$ of the base lattice $\Lambda_2(M)$ associated with $\mathcal{O}_{\mathbb{F}}$ satisfies $|\det(M)|^2=\frac{7}{4}$. Consequently,
		\[
		c_{\det}\bigl(\mathcal{C}_{\mathbb{Q}(\sqrt{-7})}\bigr)
		=
		|\gamma|\,|\alpha_1-\alpha_2|^2\,|\det(M)|^2
		=
		7.
		\]
		
		Suppose, for contradiction, that there exists another code $\widetilde{\mathcal{C}}=\mathcal{C}(\mathbb{F},\beta_1,\beta_2,\gamma)$ over $\mathcal{O}_{\mathbb{F}}$ such that $\det_{\min}(\widetilde{\mathcal{C}})=1$ and $c_{\det}(\widetilde{\mathcal{C}})<7$. Let $x^2+px+q\in\mathcal{O}_{\mathbb{F}}[x]$ be the irreducible polynomial over $\mathbb{F}$ with roots $\beta_1,\beta_2$, where $p,q\in\mathcal{O}_{\mathbb{F}}$. Then $|\beta_1-\beta_2|^2=|p^2-4q|$, and since $|\det(M)|^2=\frac{7}{4}$, the condition $c_{\det}(\widetilde{\mathcal{C}})<7$ is equivalent to
		\begin{equation}\label{eq:d7key}
			|\gamma|\,|p^2-4q|<4.
		\end{equation}
		Since $\gamma\in\mathcal{O}_{\mathbb{F}}\setminus\{0\}$ and $\mathbb{F}$ is an imaginary quadratic field, we have $|\gamma|^2=|N_{\mathbb{F}/\mathbb{Q}}(\gamma)|\in \mathbb{Z}_{\geq 1},$ and hence $|\gamma|\ge 1$. Therefore
		\begin{equation}\label{eq:d7Delta}
			|p^2-4q|<4.
		\end{equation}
		By Proposition~\ref{cases}, after translating the roots $\beta_i\mapsto\beta_i+p_0$, one may assume that $|p|\le \max\left\{1,\left|\frac{1+\sqrt{-7}}{2}\right|\right\}=\sqrt{2}$. Writing $p=a+b\omega$ with $a,b\in\mathbb{Z}$, this implies $|p|^2=a^2+ab+2b^2\le 2$, and therefore
		\[
		p\in\{0,\pm1,\pm\omega,\pm\bar\omega\},\qquad \bar\omega=\frac{1-\sqrt{-7}}{2}.
		\]
		\begin{itemize} 
			\item[\emph{Case 1:}] $p\in\{0,\pm1\}$.
			If $p=0$, then $p^2-4q=-4q$, and \eqref{eq:d7Delta} implies $|q|<1$, which is impossible since $q\neq 0$. If $p=\pm1$, then $p^2-4q=1-4q$, and \eqref{eq:d7Delta} yields $|1-4q|<4$. By the triangle inequality,  $|4q|\le |1-4q|+1<5$, hence $|q|<\frac{5}{4}$. Since $|q|^2=N_{\mathbb{F}/\mathbb{Q}}(q)\in\mathbb{Z}_{\ge 0}$, it follows that $q\in\{0,\pm1\}$. Irreducibility excludes $q=0$. If $q=-1$, then $|p^2-4q|=5$, contradicting \eqref{eq:d7Delta}. Thus $q=1$, and $|p^2-4q|=3$, so \eqref{eq:d7key} implies $|\gamma|<\frac{4}{3}<\sqrt{2}$. 
			Write $\gamma=x+y\omega$, where $\omega=\frac{1+\sqrt{-7}}{2},$ and $x,y\in\mathbb{Z}$. Then
			$
			|\gamma|^2=x^2+xy+2y^2.
			$
			If $|\gamma|<\sqrt{2}$, then
			$
			x^2+xy+2y^2<2.
			$
			Since this quantity is a nonnegative integer, it follows that $x^2+xy+2y^2\in\{0,1\}$.
			If $y\neq 0$, then $x^2+xy+2y^2\ge 2$, a contradiction. Hence $y=0$, and $x^2<2$, so $x\in\{0,\pm1\}$.
			Therefore $\gamma\in\{0,\pm1\}$.
			Since $\gamma\neq 0$, it follows that $\gamma=\pm1$. The non-norm condition forces $\gamma=-1$.
			Let $\mathbb{K}=\mathbb{F}(\beta)$, where $\beta$ is a root of $x^2+px+1$. Writing $\beta'=-p-\beta$, one has for all $u,v\in\mathbb{F}$
			\[
			N_{\mathbb{K}/\mathbb{F}}(u+v\beta)=u^2-puv+v^2.
			\]
			
			If $p=1$, taking $u=\omega$ and $v=1$, we obtain
			\[
			N_{\mathbb{K}/\mathbb{F}}(\omega+\beta)
			=
			\omega^2-\omega+1
			=
			(\omega-2)-\omega+1
			=
			-1.
			\]
			If $p=-1$, taking $u=-\omega$ and $v=1$, we obtain
			\[
			N_{\mathbb{K}/\mathbb{F}}(-\omega+\beta)
			=
			(-\omega)^2+(-\omega)+1
			=
			\omega^2-\omega+1
			=
			-1.
			\]
			Therefore, in both cases, $-1\in N_{\mathbb{K}/\mathbb{F}}(\mathbb{K}^{\times})$, contradicting the non-norm condition for $\gamma=-1$. Hence Case~1 cannot occur.

			\item[\emph{Case 2:}] $p\in\{\pm\omega,\pm\bar\omega\}$.
			Since replacing $p$ by $-p$ does not change $p^2$, and complex conjugation preserves $|p^2-4q|$, it suffices to consider $p=\omega$. In this case $|p|^2=2$, hence $|p^2|=2$. From \eqref{eq:d7Delta} it follows that $|4q|\le |p^2|+|p^2-4q|<6$, so $|q|<\frac{3}{2}$. As $|q|^2\in\mathbb{Z}_{\ge 0}$, one obtains $q\in\{0,\pm1,\pm\omega,\pm\bar\omega\}$, and irreducibility excludes $q=0$.
			
			We now compute $p^2-4q$ for the remaining possibilities. Since $p=\omega$ and $p^2=\omega-2$, we have
			\[
			p^2-4q=\omega-2-4q.
			\]
			Using $\bar{\omega}=1-\omega$, one obtains:
			\[
			\begin{array}{c|c|c}
				q & p^2-4q & |p^2-4q|^2\\
				\hline
				1 & \omega-6 & 32\\
				-1 & \omega+2 & 8\\
				\omega & -2-3\omega & 28\\
				-\omega & -2+5\omega & 44\\
				\bar{\omega} & -6+5\omega & 56\\
				-\bar{\omega} & 2-3\omega & 16
			\end{array}
			\]
			Thus, all cases except $q=-1$ contradict the strict inequality \eqref{eq:d7Delta}, because they give $|p^2-4q|\geq 4$.
			
			It remains to consider $q=-1$. In this case,
			\[
			|p^2-4q|=|\omega+2|=2\sqrt{2}<4,
			\]
			so the inequality \eqref{eq:d7Delta} alone does not exclude this possibility. However, if $\beta$ is a root of $
			x^2+\omega x-1,
			$
			then
			$
			N_{\mathbb{K}/\mathbb{F}}(\beta)=\beta\beta'=q=-1.
			$
			Therefore,
			$-1\in N_{\mathbb{K}/\mathbb{F}}(\mathbb{K}^{\times}),
			$
			which contradicts the non-norm condition for $\gamma=-1$.
			Consequently, Case~2 cannot occur.
			
		\end{itemize}
		Thus, every possible reduced representative of $p$ leads to a
		contradiction under the assumption $c_{\det}(\widetilde{\mathcal{C}})<7$.
		Therefore, no admissible code can have determinant-based metric smaller
		than $7$, and the asserted optimality follows.
	\end{proof}

	\subsection{Optimal $2\times 2$ Space-Time Block Codes over $\mathbb{Z}[\frac{1+\sqrt{-11}}{2}]$}

	Finally, we consider $\mathbb{F}=\mathbb{Q}(\sqrt{-11})$ and determine the corresponding optimal construction.
	By Proposition~\ref{-1padic}, the element $-1$ is not a norm from $\mathbb{K}=\mathbb{Q}(\sqrt{-11},\sqrt{-3})$ to $\mathbb{F}$, and we therefore take $\gamma=-1$.

	\begin{theorem}\label{thm:d11}
		The code $\mathcal{C}_{\mathbb{Q}(\sqrt{-11})}=\mathcal{C}(\mathbb{F},\alpha_1,\alpha_2,\gamma)$, with $\mathbb{F}=\mathbb{Q}(\sqrt{-11})$, $\alpha_1=\frac{1+\sqrt{-3}}{2},\,\alpha_2=\frac{1-\sqrt{-3}}{2}$ and $\gamma=-1$ as above, is optimal among all $2\times 2$ space–time block codes defined over $\mathbb{Z}\!\left[\frac{1+\sqrt{-11}}{2}\right]$ with minimum determinant equal to $1$.
	\end{theorem}
	\begin{proof} 
		By Lemma~\ref{dmin1}, all codes under consideration have minimum determinant equal to $1$, so optimality is determined by the determinant-based metric $c_{\det}$.
		For the reference code, one has $\alpha_1-\alpha_2=\sqrt{-3}$, hence $|\alpha_1-\alpha_2|^2=3$, and $\gamma=-1$, so $|\gamma|=1$. Since $11\equiv 3 \pmod{4}$, the generator matrix $M$ of the base lattice $\Lambda_2(M)$ associated with $\mathcal{O}_{\mathbb{F}}$ satisfies $|\det(M)|^2=\frac{11}{4}$. Therefore,
		\[
		c_{\det}\bigl(\mathcal{C}_{\mathbb{Q}(\sqrt{-11})}\bigr)
		=
		|\gamma|\,|\alpha_1-\alpha_2|^2\,|\det(M)|^2
		=
		\frac{33}{4}.
		\]
		
		Suppose, for contradiction, that there exists another code $\widetilde{\mathcal{C}}=\mathcal{C}(\mathbb{F},\beta_1,\beta_2,\gamma)$ over $\mathcal{O}_{\mathbb{F}}$ such that $\det_{\min}(\widetilde{\mathcal{C}})=1$ and $c_{\det}(\widetilde{\mathcal{C}})<\frac{33}{4}$. Let $x^2+px+q\in\mathcal{O}_{\mathbb{F}}[x]$ be the irreducible polynomial over $\mathbb{F}$, and let $\beta_1,\beta_2$ denote its roots, where $p,q\in\mathcal{O}_{\mathbb{F}}$. Then $|\beta_1-\beta_2|^2=|p^2-4q|$. Since $|\det(M)|^2=\frac{11}{4}$, the inequality $c_{\det}(\widetilde{\mathcal{C}})<\frac{33}{4}$ is equivalent to
		\begin{equation}\label{eq:d11key}
			|\gamma|\,|p^2-4q|<3.
		\end{equation}
		Since $\gamma\in\mathcal{O}_{\mathbb{F}}
		\setminus\{0\}$ and $\mathbb{F}$ is an imaginary quadratic field, we  have $|\gamma|^2\in \mathbb{Z}_{\geq 1},$ and hence $|\gamma|\geq 1$. Therefore
		\begin{equation}\label{eq:d11Delta}
			|p^2-4q|<3.
		\end{equation}
		By Proposition~\ref{cases}, after a suitable translation of the roots $\beta_i\mapsto \beta_i+p_0$, we may assume that $|p|\le \max\left\{1,\left|\frac{1+\sqrt{-11}}{2}\right|\right\}=\sqrt{3}$. Writing $p=a+b\omega$, where $\omega=\frac{1+\sqrt{-11}}{2}$ and $a,b\in\mathbb{Z}$, we obtain $|p|^2=a^2+ab+3b^2\le 3$. Consequently,
		$
		p\in\{0,\pm1,\pm\omega,\pm\bar\omega\},$ where $\bar\omega=\frac{1-\sqrt{-11}}{2}.
		$
		\begin{itemize} 
			\item[\emph{Case 1:}] $p=0$.
			In this case $p^2-4q=-4q$, and \eqref{eq:d11Delta} implies $4|q|<3$, hence $|q|<\frac{3}{4}<1$. This is impossible, since $q\neq 0$ by irreducibility and $|q|^2=N_{\mathbb{F}/\mathbb{Q}}(q)\in\mathbb{Z}_{\ge 1}$ for every nonzero $q\in\mathcal{O}_{\mathbb{F}}$.
			\item[\emph{Case 2:}] $p=\pm1$.
			Here $p^2-4q=1-4q$, and \eqref{eq:d11Delta} gives $|1-4q|<3$. The triangle inequality yields $|4q|\le |1-4q|+1<4$, hence $|q|<1$, which forces $q=0$, contradicting irreducibility.
			\item[\emph{Case 3:}] $p\in\{\pm\omega,\pm\bar\omega\}$.
			Since replacing $p$ by $-p$ does not change $p^2$, and conjugation preserves the modulus of $p^2-4q$, it suffices to consider $p=\omega$. As $|p|^2=|\omega|^2=N_{\mathbb{F}/\mathbb{Q}}(\omega)=3$, one has $|p^2|=3$. From \eqref{eq:d11Delta},
			\[
			|4q|=|p^2-(p^2-4q)|\le |p^2|+|p^2-4q|<6,
			\]
			so $|q|<\frac{3}{2}$. Hence $|q|^2=N_{\mathbb{F}/\mathbb{Q}}(q)\in\{0,1,2\}$, and since $q\neq 0$, it follows that $N_{\mathbb{F}/\mathbb{Q}}(q)\in\{1,2\}$.
			The norm form on $\mathcal{O}_{\mathbb{F}}=\mathbb{Z}[\omega]$ is $N_{\mathbb{F}/\mathbb{Q}}(a+b\omega)=a^2+ab+3b^2$. A straightforward check  shows that the equation $a^2+ab+3b^2=2$ has no integer solution. Therefore $N_{\mathbb{F}/\mathbb{Q}}(q)=1$, and thus $q=\pm1$.\\
			\textbf{Subcase 3.1:} $q=1$.
			Using $\omega^2=\omega-3$, one obtains $p^2-4q=\omega^2-4=\frac{-13+\sqrt{-11}}{2}$. Hence $N_{\mathbb{F}/\mathbb{Q}}(p^2-4q)=\frac{(-13)^2+11}{4}=45$, so $|p^2-4q|=\sqrt{45}>3$, contradicting \eqref{eq:d11Delta}.\\
			\textbf{Subcase 3.2:} $q=-1$.
			Again using $\omega^2=\omega-3$, one obtains $p^2-4q=\omega^2+4=\frac{3+\sqrt{-11}}{2}$. Thus $N_{\mathbb{F}/\mathbb{Q}}(p^2-4q)=\frac{3^2+11}{4}=5$, and $|p^2-4q|=\sqrt{5}$. From \eqref{eq:d11key} it follows that $|\gamma|<\frac{3}{\sqrt{5}}<\sqrt{3}$.
			Write $\gamma=x+y\omega$, where $\omega=\frac{1+\sqrt{-11}}{2}$ and $x,y\in\mathbb{Z}$. Then
			$
			|\gamma|^2=x^2+xy+3y^2.
			$
			If $|\gamma|<\sqrt{3}$, then
			$
			x^2+xy+3y^2<3.
			$
			Since this quantity is a nonnegative integer, it follows that
			$
			x^2+xy+3y^2\in\{0,1,2\}.
			$
			If $y\neq 0$, then $3y^2\ge 3$, a contradiction. Hence $y=0$, and so $x^2<3$, which gives $x\in\{0,\pm1\}$.
			Therefore $\gamma\in\{0,\pm1\}$.
			Since $\gamma\neq 0$, it follows that $\gamma=\pm1$. As $1=N_{\mathbb{K}/\mathbb{F}}(1),$
			the non-norm condition excludes $\gamma=1,$ and therefore $\gamma=-1$.
			Let $\beta$ be a root of $x^2+\omega x-1$. Then $N_{\mathbb{K}/\mathbb{F}}(\beta)=\beta\beta'=q=-1$, so $-1$ is a norm from $\mathbb{K}=\mathbb{F}(\beta)$ to $\mathbb{F}$, contradicting the non-norm condition.
		\end{itemize} Thus, every possible value of $p$ leads to a contradiction. Therefore, no code $\widetilde{\mathcal{C}}$ satisfies $c_{\det}(\widetilde{\mathcal{C}})<\frac{33}{4}$, and the optimality of the proposed construction follows.
		
	\end{proof}

	\subsection{Global Optimality over Imaginary Quadratic Integer Rings}

	From the analysis carried out in the previous subsections, it follows that for each imaginary quadratic field $\mathbb{F} = \mathbb{Q}(\sqrt{-d})$, with $d \in \{2,3,7,11\}$, the corresponding optimal code $\mathcal{C}$ satisfies
	$
	c_{\det}(\mathcal{C}) \geq c_{\det}(\mathcal{C}_{\mathbb{Q}(\sqrt{-3})}).
	$
	In particular, no code defined over $\mathcal{O}_{\mathbb{F}}$ achieves a strictly smaller determinant-based metric than the Eisenstein construction.
	
	This leads to the following global optimality result.
	
	\begin{theorem}
		The Eisenstein code
		\[
		\mathcal{C}_{\mathbb{Q}(\sqrt{-3})}
		= \mathcal{C}\!\left(\mathbb{Q}(\sqrt{-3}),
		\frac{-p + \sqrt{p^2 - 4q}}{2},
		\frac{-p - \sqrt{p^2 - 4q}}{2},
		\zeta_{6}
		\right),
		\]
		where $p = 1+ \zeta_6$ and $q = \sqrt{-3}$, is optimal among all $2 \times 2$ space-time block codes with minimum determinant equal to $1$ defined over the rings of integers $\mathcal{O}_{\mathbb{F}}$, with $\mathbb{F} = \mathbb{Q}(\sqrt{-d})$ and $d$ a positive square-free integer.
	\end{theorem}
	
	\begin{proof}
		The optimality of the Eisenstein construction over $\mathbb Q(\sqrt{-3})$ within the corresponding quadratic-field family follows from \cite{wang}. 
		Directly by Theorems 5.1, 5.2, and 5.3, the optimal codes over the fields $\mathbb{Q}(\sqrt{-d})$ for $d \in \{2,7,11\}$ have determinant-based metrics strictly greater than or equal to that of the Eisenstein code.  Moreover, the reduction argument established earlier shows that no field outside the finite set $d \in \{1,2,3,7,11\}$ can provide a better construction.  Since the Gaussian case $d=1$ is already known to be suboptimal compared to the Eisenstein case \cite{wang}, it follows that the code over $\mathbb{Q}(\sqrt{-3})$ minimizes the determinant-based metric globally.
		
		Therefore, the Eisenstein construction achieves the maximum normalized density among all $2 \times 2$ space-time block codes arising from imaginary quadratic integer rings.
	\end{proof}

	The optimal codes obtained in this work in the previous sections, together with those presented in \cite{wang}, are summarized in  Table~1.
	The constructions over $\mathbb{Q}(\sqrt{-2})$ and $\mathbb{Q}(\sqrt{-7})$ arise from the quaternion division algebras $(-3,-1)_{\mathbb{Q}(\sqrt{-2})}$ and $(-1,-1)_{\mathbb{Q}(\sqrt{-7})}$, respectively, and yield realizations of the $E_8$ lattice, as described in \cite{carina}. In addition,   for $\mathbb{F}=\mathbb{Q}(\sqrt{-7})$, the associated quaternion algebra $(-1,-1)_{\mathbb{F}}$ gives rise to the well-known Silver code, which achieves normalized minimum determinant $\delta(\Lambda)=\rho(\Lambda)^{\frac{1}{4}}=1/\sqrt{7}$ (see \cite{silver}).

	This completes the classification of optimal $2\times 2$ space-time block codes within the considered family.

	\begin{table}[ht!] \label{tab1} 
		\tbl{Optimal $2\times 2$ STBCs over the rings of integers of  imaginary quadratic fields
			$\mathbb{F}=\mathbb{Q}(\sqrt{-d})$. The table lists the corresponding
			quadratic extension, defining polynomial, quaternion algebra, and
			normalized density.}
		{\begin{tabular}{@{}ccccc@{}} \toprule
				$\mathbb{F}$ 
				& $\mathbb{K} = \mathbb{F}(\sqrt{p^2-4q})$ 
				& $f(x)$ 
				& $(p^2-4q,\gamma)_{\mathbb{F}}$ 
				& $\rho(\Lambda)$ \\
				\colrule
				
				$\mathbb{Q}(\sqrt{-1})$ 
				& $\mathbb{F}(\sqrt{3})$ 
				& $x^2 + \sqrt{-1}x - 1$ 
				& $(3,\,1+\sqrt{-1})_{\mathbb{F}}$ 
				& $\frac{1}{18}\approx 0.0556$ \\
				
				$\mathbb{Q}(\sqrt{-2})$ 
				& $\mathbb{F}(\sqrt{-3})$ 
				& $x^2 - x + 1$ 
				& $(-3,\,-1)_{\mathbb{F}}$ 
				& $\frac{1}{36}\approx 0.0278$ \\
				
				$\mathbb{Q}(\sqrt{-3})$ 
				& $\mathbb{F}(\sqrt{(1+\zeta_6)^2+4\sqrt{-3}})$ 
				& $x^2 + (1+\zeta_6)x + \sqrt{-3}$ 
				& $((1+\zeta_6)^2 - 4\sqrt{-3},\,\zeta_{6})_{\mathbb{F}}$ 
				& $\frac{16}{189}\approx 0.0847$ \\
				
				$\mathbb{Q}(\sqrt{-7})$ 
				& $\mathbb{F}(\sqrt{-1})$ 
				& $x^2 + 1$ 
				& $(-1,\,-1)_{\mathbb{F}}$ 
				& $\frac{1}{49}\approx 0.0204$ \\
				
				$\mathbb{Q}(\sqrt{-11})$ 
				& $\mathbb{F}(\sqrt{-3})$ 
				& $x^2 - x + 1$ 
				& $(-3,\,-1)_{\mathbb{F}}$ 
				& $\frac{16}{1089}\approx 0.0147$ \\
				
				\botrule
		\end{tabular}}
	\end{table}

	\section{Conclusion}
	
	In this paper, using a determinant-based comparison criterion, we determine optimal $2\times 2$ space-time block codes over rings of integers of $\mathbb{Q}(\sqrt{-d})$, and reduce the problem to the finite set $d \in \{2,7,11\}$. We prove that the Eisenstein construction remains optimal among all such codes, in the sense that it achieves the largest normalized density.
	For each case, constructions are provided and their determinant-based metrics are computed. In particular, for $d=7$, the resulting normalized minimum determinant coincides with
	that of the Silver code. The constructions over $\mathbb{Q}(\sqrt{-2})$ and $\mathbb{Q}(\sqrt{-7})$ arise from quaternion division algebras and lead to realizations of the $E_8$ lattice.
	A key ingredient is the use of non-norm criteria for quadratic extensions.  These criteria ensure the division algebra property. Although decoding algorithms are not the main focus of this work, the constructions considered here are compatible with algebraic reduction techniques over imaginary quadratic fields, such as the LLL-type reductions, which extend decoding techniques beyond the Gaussian and Eisenstein settings.
	Future work includes extensions to higher dimensions and a study of the associated lattices arising from maximal orders in cyclic division algebras.\\

	\noindent \textbf{Acknowledgements:}
	This work was partially supported by CNPq under Grant No. 405842/2023-6 and by FAPESP under grants 2019/20800-8, 2022/02303-0,  2023/15735-8,  2024/00923-6.

\end{document}